\documentclass[aps,prb,reprint,superscriptaddress]{revtex4-1}

\usepackage{graphicx}
\usepackage[english]{babel}
\usepackage{amsmath}
\usepackage{SIunits}

\begin{document}

\title{Dry transfer of CVD graphene using MoS$_2$-based stamps}

\author{Luca Banszerus}
\affiliation{JARA-FIT and 2nd Institute of Physics, RWTH Aachen University, 52074 Aachen, Germany}
\affiliation{Peter Gr\"unberg Institute (PGI-9), Forschungszentrum J\"ulich, 52425 J\"ulich, Germany}
\author{Kenji Watanabe}
\author{Takashi Taniguchi}
\affiliation{National Institute for Materials Science, 1-1 Namiki, Tsukuba, 305-0044, Japan}
\author{Bernd Beschoten}
\affiliation{JARA-FIT and 2nd Institute of Physics, RWTH Aachen University, 52074 Aachen, Germany}
\author{Christoph Stampfer}
\thanks{E-mail address: stampfer@physik.rwth-aachen.de}
\affiliation{JARA-FIT and 2nd Institute of Physics, RWTH Aachen University, 52074 Aachen, Germany}
\affiliation{Peter Gr\"unberg Institute (PGI-9), Forschungszentrum J\"ulich, 52425 J\"ulich, Germany}

\date{\today}

\begin{abstract}
Recently, a contamination-free dry transfer method for graphene grown by chemical vapor deposition (CVD) has been reported that allows to directly pick-up graphene from the copper growth substrate using a flake of hexagonal boron nitride (hBN), resulting in ultrahigh charge carrier mobility and low overall doping. Here, we report that not only hBN, but also flakes of molybdenum disulfide (MoS$_2$) can be used to dry transfer graphene. This, on one hand, allows for the fabrication of complex van-der-Waals heterostructures using CVD graphene combined with different two-dimensional materials and, on the other hand, can be a route towards a scalable dry transfer of CVD graphene. The resulting heterostructures are studied using low temperature transport measurements revealing a strong charge carrier density dependence of the carrier mobilities (up to values of 12,000~cm$^2$/(Vs)) and the residual charge carrier density fluctuations near the charge neutrality point when changing the carrier density in the MoS$_2$ by applying a top gate voltage.
\end{abstract}

\maketitle

The high room temperature mobility\cite{Bol08,Wan13,Ban16} and the tunable charge carrier density make graphene an interesting material for many applications such as high frequency electronics\cite{Lin10}, ultra-sensitive Hall sensors\cite{Dau15,Wan16} and spintronics\cite{Wei14,Dro16}. In order to realize such applications, it is necessary to make high quality graphene available on a large scale. Graphene grown by chemical vapor deposition (CVD) has recently made numerous advances concerning its growth\cite{Che13,Li09,Bae10,Li11} and transfer\cite{Ban15,Suk11,Pet12,Piz15}. We previously reported that the electronic properties of CVD graphene are equivalent to devices built from high quality exfoliated graphene if transfer-related degradations and contaminations are avoided\cite{Ban15,Ban16}. The highest electronic quality in CVD graphene has so far been achieved by using exfoliated hexagonal boron nitride (hBN) crystals by (1) picking-up CVD-graphene directly from the catalytic copper foil (substrate material) and by (2) subsequently encapsulating it with another hBN crystal\cite{Ban16}. Here, we report on CVD-graphene that has been dry-transferred from the copper foil using a similar scheme. Instead of hBN, we use molybdenum disulfide (MoS$_2$) to transfer graphene. Expanding this transfer process from using flakes of exfoliated hexagonal boron nitride to a larger class of two-dimensional (2d) materials has numerous advantages: Firstly, van-der-Waals heterostructures consisting of different 2d materials have attracted large attention in recent years, as they allow for new device properties, e.g. proximity induced spin-orbit interaction\cite{Wan15,Avs14} or applications in the field of optoelectronics\cite{Brit13}. Secondly, high quality large area hBN with a low adhesion to its substrate has not been successfully grown so far, which limits the size of the heterostructures that can be obtained using the dry transfer to the size of the exfoliated hBN flake. Thus, finding alternative, scalable 2d materials to transfer graphene and to serve as a substrate that preserves the intrinsic electronic properties of graphene could speed up the scaling, opening up the way towards true high quality graphene applications. Transition metal dichalcogenides (TMDCs) such as MoS$_2$ can by now be grown on different substrate materials\cite{Las13,Eic15,Che16} such as sapphire with high structural and electronic quality. Besides opening up a larger set of possible material combinations to enable new device functionalities, using a broader set of synthetic and thus potentially scalable 2d materials for the transfer could be a future route towards scaling high quality CVD graphene to arbitrary sizes. Our findings suggests that, similar to the established stacking techniques for exfoliated van-der-Waals materials\cite{Wan13} a much wider range of 2d materials can be used for the transfer process.

Graphene is grown using a low pressure CVD process on the inside of enclosures folded from copper foil\cite{Che13}, resulting in individual graphene crystals of a few hundred micrometer in size on the copper. In order to weaken the adhesion between the graphene and the copper substrate and thus facilitate the transfer process, the graphene is stored under ambient conditions for a few days, during which a thin cuprous oxide (Cu$_2$O) layer forms at the graphene-to-Cu interface\cite{Ban15,Lu12}. An optical image of a typical graphene crystal with an oxidized interface is shown in Fig.~1a. Following our previous reports on dry graphene transfer\cite{Ban15,Ban16}, a polymer stack consisting of a thick layer of poly(vinyl alcohol) (PVA) and a thin layer of poly(methyl methacrylate) (PMMA) is prepared. After exfoliating MoS$_2$ flakes of various thicknesses between 10~nm and 70~nm on the polymer, the stack is placed on a polydimethylsiloxane (PDMS) stamp. Using a mask aligner, the TMDC is brought into contact with the graphene at 125~$^{\circ}$C. After picking-up the graphene, the MoS$_2$/graphene stack is placed on an exfoliated hBN flake. Thereafter, the polymers are dissolved in water, acetone and isopropanol. Fig.~1b shows an optical microscope image of a heterostructure consisting of hBN, graphene and MoS$_2$.\\

\begin{figure}[tb]
\centering
\includegraphics[draft=false,keepaspectratio=true,clip,width=\linewidth]{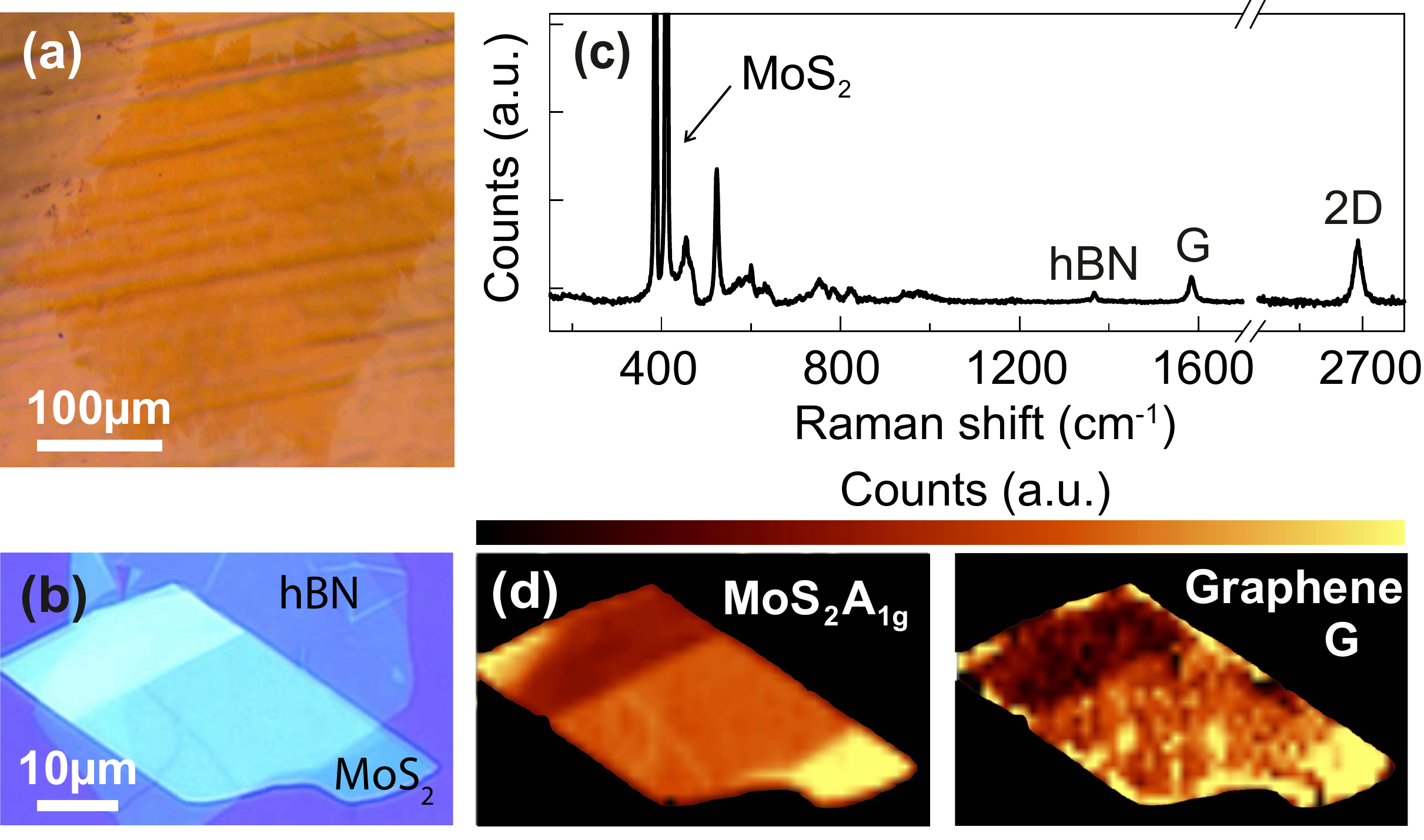}
\caption[Fig01]{\textbf{(a)} Optical image of a CVD grown graphene flake on copper foil. \textbf{(b)} Microscope image of a dry transferred hetero-stack consisting of hBN/graphene/MoS$_2$. \textbf{(c)} Typical Raman spectrum of graphene encapsulated between hBN and MoS$_2$. \textbf{(d)} left: Raman map of the intensity of the MoS$_2$ A$_{\mathrm{1g}}$ peak, measured for the flake depicted in (b); right: Raman map of the intensity of the graphene G-peak. }
\label{f1}
\end{figure}

 We use scanning confocal Raman microscopy which is a fast and non-invasive optical method to probe the structural and electronic properties of graphene including defects, doping and strain, as well as nm-scale strain variations\cite{Fer06,Gra07,For13,Neu14,Lee12}. A typical Raman spectrum of a MoS$_2$/graphene/hBN heterostructure is shown in Fig.~1c. The E$^1_{\mathrm{2g}}$ and the  A$_{\mathrm{1g}}$ mode of MoS$_2$ are centered at 386~cm$^{-1}$ and 411~cm$^{-1}$, respectively\cite{Ton13,Sah13}. The hBN peak is centred at 1365~cm$^{-1}$. The graphene G-peak is located around 1582~cm$^{-1}$ and the 2D-peak is centred at 2686~cm$^{-1}$ indicating low doping and little strain in the transferred graphene layer\cite{Lee12}. Compared to graphene encapsulated between two flakes of hBN, the full-width-at-half-maximum (FWHM) of the 2D peak, $\Gamma_\mathrm{2D}$, is slightly elevated to around 20~cm$^{-1}$ indicating still low amounts of nanometre-scale strain variations within the laser spot\cite{Neu14}, which is a good indication for high charge carrier mobility in the graphene\cite{Cuo14}. A more detailed study on strain and doping inhomogeneities of graphene on MoS$_2$ and other substrate materials has recently been published\cite{Ban17}. These findings are very similar to those obtained for high quality graphene transferred using hBN\cite{Ban15,Ban16}. Fig.~1d shows the Raman maps of the A$_{\mathrm{1g}}$ mode of MoS$_2$ and the intensity of the graphene G-peak, corresponding to the heterostructure depicted in Fig.~1b. The data shows that the entire MoS$_2$ flake is covered with graphene, demonstrating a reliable transfer process. \\

In order to investigate the charge transport properties of the resulting van-der-Waals stack, we fabricated dual-gated Hall bar devices with one-dimensional Cr/Au edge contacts (see left inset of Fig.~2b). The Hall bar is patterned from the heterostructure by electron beam lithography and reactive ion etching using argon and SF$_6$ as etching gases. Contacts are fabricated by electron beam lithography followed by electron beam evaporation of Cr and Au. Fig.~2a shows the four-terminal sheet resistivity of the device as function of the applied top gate voltage, $V_\mathrm{TG}$, and the applied back gate voltage, $V_\mathrm{BG}$, measured at a temperature of 1.6~K. Fig.~2b depicts line cuts through the data at V$_\mathrm{TG}=3$~V and $V_\mathrm{TG}=-1$~V (blue and red line, respectively). For top gate voltages above 2.8~V, a clear and sharp resistance peak is observed (see blue line in Fig.~2b and upper part of Fig.~2a), where the graphene is screened from the top gate, as seen by the constant position of the resistance peak when changing V$_\mathrm{TG}$ (red area in the upper part of Fig.~2a). This indicates that the Fermi energy in the MoS$_2$ is tuned into its conduction band, allowing charge carriers in the MoS$_2$ to screen the applied top gate potential (see right inset of Fig.~2b). In contrast, for $V_\mathrm{TG}~<~2$~V, the Fermi energy of MoS$_2$ has moved into the band gap, allowing to continuously tune the charge carrier density in graphene by changing the top gate voltage. At the same time, the absence of charge carriers in the MoS$_2$ leads to a decreased dielectric screening of charge traps and defects located in MoS$_2$ and the MoS$_2$/graphene interface, which strongly increase the residual charge carrier density fluctuations in graphene. This results in a reduced maximum resistance and a broadening of the resistance peak as seen for the red trace in Fig.~2b and the lower part of Fig.~2a at the charge neutrality point.\\

\begin{figure}[tb]
\centering
\includegraphics[draft=false,keepaspectratio=true,clip,width=0.9\linewidth]{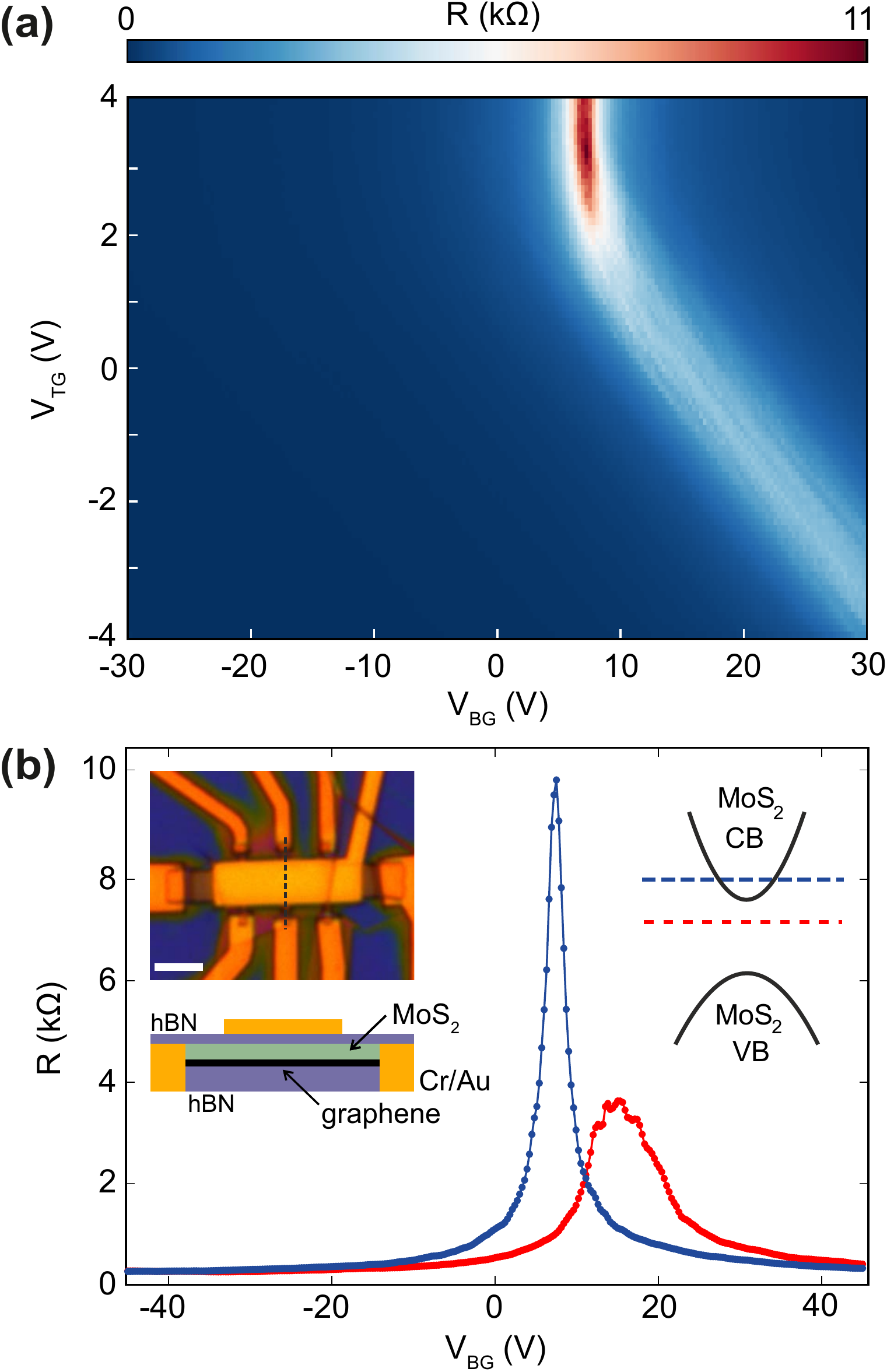}
\caption[Fig02]{\textbf{(a)} Resistivity of the MoS$_2$/graphene/hBN Hall bar as function of the applied top gate and back gate voltage, measured at 1.6~K. \textbf{(b)} Line cuts of the data shown in (a), taken at $V_{\mathrm{TG}}~=~3$~V (blue) and $V_{\mathrm{TG}}~ =~ -1$~V (red). Inset left: optical microscope image of the device. The scale bar is 4~$\mu$m. Below, a schematic cross-section of the sample is depicted. Inset right: Schematic band structure of the MoS$_2$. The top gate allows to either tune the Fermi energy into the band gap or into the conduction band. }
\label{f2}
\end{figure}

\begin{figure}[tb]
\centering
\includegraphics[draft=false,keepaspectratio=true,clip,width=1.0\linewidth]{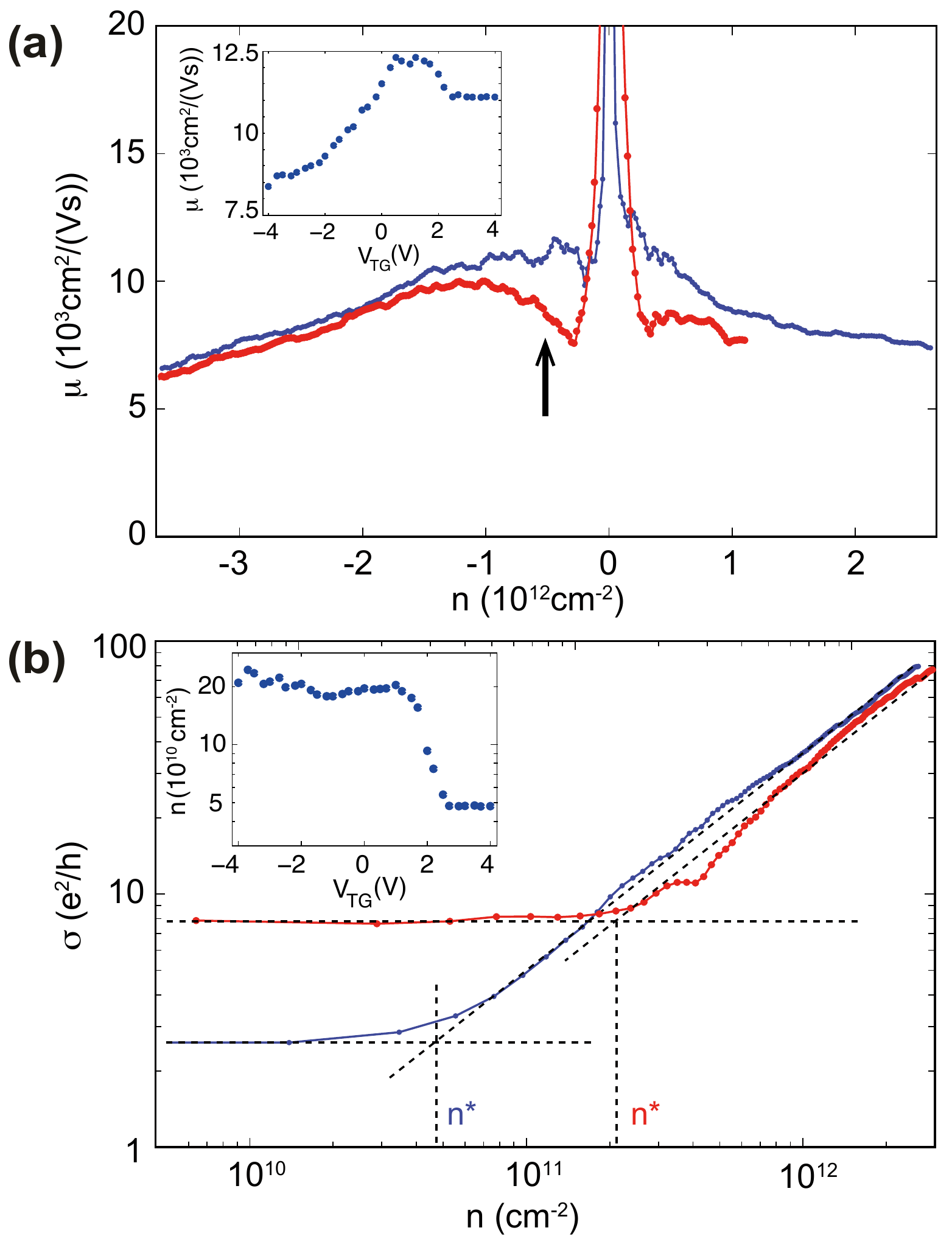}
\caption[Fig03]{\textbf{(a)} Charge carrier mobility $\mu$ vs. charge carrier density $n$ for two back gate traces in Fig.~2a at $V_{\mathrm{TG}}~=~3$~V (blue) and $V_{\mathrm{TG}}~=~-3$~V (red) taken at $T=1.6$~K. The inset depicts the mobility at a fixed density of $n~=~-0.5\times 10^{12}$~cm$^{-2}$ (indicated by the black arrow) as function of the top gate voltage. \textbf{(b)} Conductance $\sigma$ vs. charge carrier density $n$ for the measurements in (a) plotted on a double logarithmic scale which allows to extract the residual charge carrier density fluctuations $n^*$ near at the charge neutrality point. The inset presents the top gate dependence of $n^*$.}
\label{f2}
\end{figure}

For quantifying the influence of the disorder potential in the MoS$_2$ on the charge transport in the graphene layer, in both the screened $V_\mathrm{TG}~>~2.8$~V and unscreened $V_\mathrm{TG}~<~2.8$~V case, we extract the charge carrier mobility $\mu$ and the residual charge carrier density fluctuations $n^*$ at the charge neutrality point. Fig.~3a depicts the field-effect mobility determined by the Drude formula $\sigma=ne\mu$ as function of the charge carrier density in the graphene layer at constant top gate voltages of  $V_{\mathrm{TG}}$~=~3~V (blue) and at $V_{\mathrm{TG}}~=~-3$~V (red). The extracted mobility is in both cases on the order of $\mu~=~$10,000~cm$^2$/(Vs). At high charge carrier densities, the mobility is independent of whether or not the Fermi level of MoS$_2$ is tuned into its conduction band. We note that the MoS$_2$ is not significantly contributing to transport, as the mobility in MoS$_2$ is typically orders of magnitudes lower than in graphene. More importantly, the formation of a Schottky barrier further suppresses transport through the MoS$_2$\cite{ref}. Evidence for the absence of a significant parallel conducting channel through the MoS$_2$ can be seen in Fig.~2a. If the MoS$_2$ was contributing significantly to transport, an increase of the conductivity is expected, once there are free carriers in the MoS$_2$ layer, i.e. at $V_\mathrm{TG}=2.8V$, which is not observed in the experiment. Similar observations have been made in previous reports of graphene on other TMDC materials\cite{Wan15,Avs14}. The inset of Fig.~3a presents the top gate dependence of the charge carrier mobility at a constant charge carrier density of $n~=-~0.5\times 10^{12}$~cm$^{-2}$ in the graphene layer. The mobility of the graphene increases from $\mu~=~$8,000~cm$^2$/(Vs) to around $\mu~=~$11,000~cm$^2$/(Vs) when populating the conduction band of the MoS$_2$. We attribute this behaviour to the self-screening of charge carriers in the graphene from the disorder potential in the MoS$_2$. However, at low charge carrier densities, the extracted mobility decreases for $V_{\mathrm{TG}}~=~-3$~V (red curve in Fig.~3a), due to more pronounced electron-hole puddles and potential charge transfer into trap states located at the MoS$_2$ interface. These effects are less present, when the Fermi energy is located in the conduction band of the MoS$_2$ as charge carriers can screen the Coulomb potential and charge traps are already occupied by carriers in the MoS$_2$ (blue curve in Fig.~3a).

We now focus on the charge carrier density fluctuations near the charge neutrality point by plotting the conductance vs. charge carrier density on a double logarithmic scale (Fig.~3b) for both traces shown in Fig.~3a. Following the scheme of Couto \textit{et al.}\cite{Cuo14}, we perform line fits to this double logarithmic representation of the data in order to extract $n^*$. At $V_{\mathrm{TG}}~=~ 3$~V, where the Fermi energy of MoS$_2$ is tuned into its conduction band, we extract $n^*~=~ 4.8\times 10^{10}$~cm$^{-2}$, while in the case where the Fermi energy lies in the band gap, we measure $n^*~=~ 2.2\times 10^{11}$~cm$^{-2}$. This drastic increase of the charge carrier density fluctuations at charge neutrality by almost one order of magnitude (See also inset of Fig.~3b) demonstrates the importance of a homogeneously charged substrate and the absence of charge traps for high quality graphene devices. Furthermore, we emphasize that the concentration of Coulomb scatterers and defects in the MoS$_2$ is subject to growth methods and fabrication techniques and might be heavily improved by processing in a glove box and direct encapsulation with hBN. Furthermore, the presence of a band gap in MoS$_2$ allows to precisely tune the number of charge carriers, available for screening in the substrate material, which might be of use, for example, when studying the interaction between two van-der-Waals materials.  \\

In this work, we demonstrated that MoS$_2$ crystals can be used to delaminate CVD graphene from the underlying copper showing that the dry transfer method can potentially be applied to a large number of other 2d materials resulting in more complex van-der-Waals heterostructures. This allows for tailoring the electronic properties of the resulting heterostructure by combining appropriate combinations of 2d materials, as has been demonstrated previously in heterostructures assembles from exfoliated flakes. Confocal Raman microscopy verifies the high structural quality, reflected in a low values of $\Gamma_\mathrm{2D}$. Low temperature transport measurements show carrier mobilities on the order of $\mu~=~$10,000~cm$^2$/(Vs), which are lower than what has been reported for dry transferred CVD graphene encapsulated in hBN. We attribute this observation to scattering with a strongly varying disorder potential and charge transfer into trap states present in the MoS$_2$. We demonstrate that both, the charge carrier mobility, as well as the charge carrier density fluctuations at the charge neutrality point of the graphene are affected by the disorder potential and charge traps. By increasing the charge carrier density in the MoS$_2$ by a top gate voltage, its scattering potential can be screened, allowing to tune the electronic properties of the graphene.\\

\begin{acknowledgements}
Work supported by  the  EU  project  Graphene  Flagship (contract no.  696656), the ERC (contract no. 280140), the DFG (SPP-1459, BE 2441/9-1). Growth of hexagonal boron nitride crystals was supported by the Elemental Strategy Initiative conducted by the MEXT, Japan and JSPS KAKENHI Grant Numbers JP26248061, JP15K21722 and JP25106006.
\end{acknowledgements}

%

\begin{thebibliography}{[1]}

\bibitem{Bol08}
K.I. Bolotin, K.J. Sikes, Z. Jiang, M. Klima, G. Fudenberg, J. Hone, P. Kim, H.L. Stormer,
Ultrahigh electron mobility in suspended graphene. Solid State Commun. \textbf{146}, 351, (2011).


\bibitem{Wan13}
L. Wang, I. Meric, P. Y. Huang, Q. Gao, Y. Gao, H. Tran, T. Taniguchi, K. Watanabe, L. M. Campos, D. A. Muller, J. Guo, P. Kim, J. Hone, K. L. Shepard, and C. R. Dean, One-Dimensional Electrical Contact to a Two-Dimensional Material. Science \textbf{342}, 614 (2013).

\bibitem{Ban16}
L. Banszerus, M. Schmitz, S. Engels, M. Goldsche, K. Watanabe, T. Taniguchi, B. Beschoten, and C. Stampfer, Ballistic Transport Exceeding 28 $\mu$m in CVD Grown Graphene. Nano Lett. \textbf{16}, 1387-1391, (2016).

\bibitem{Lin10}
Y. Lin, C. Dimitrakopoulos, K. Jenkins, D. Farmer, H. Chiu, A. Grill and P. Avouris, 100-GHz transistors from wafer-scale epitaxial graphene. Science \textbf{327}, 662-662, (2010).


 \bibitem{Dau15}
 J. Dauber, A. A. Sagade, M. Oellers, K. Watanabe, T. Taniguchi, D. Neumaier, and C. Stampfer, Ultra-sensitive Hall sensors based on graphene encapsulated in hexagonal boron nitride.  Appl. Phys. Lett. \textbf{106}, 193501 (2015).


\bibitem{Wan16}
 Z. Wang, L. Banszerus, M. Otto, K. Watanabe, T. Taniguchi, C. Stampfer and D. Neumaier, Encapsulated graphene-based Hall sensors on foil with increased sensitivity. Phys. Status Solidi B., doi: 10.1002/pssb.201600224, (2016).

\bibitem{Wei14}
H. Wei, R.K. Kawakami, M. Gmitra, and J. Fabian, Graphene spintronics. Nature Nanotechnol. \textbf{9}, 794, (2014).

\bibitem{Dro16}
 M. Dr\"ogeler, C. Franzen, F. Volmer, T. Pohlmann, L. Banszerus, M. Wolter, K. Watanabe, T. Taniguchi, C. Stampfer, and B. Beschoten,  Spin Lifetimes Exceeding 12 ns in Graphene Nonlocal Spin Valve Devices. Nano Lett. \textbf{16}, 3533-3539, (2016).






\bibitem{Che13}
S. Chen, H. Ji, H. Chou, Q. Li, H. Li, J. W. Suk, R. Piner, L. Liao, W. Cai, R. S. Ruoff, Millimeter-size single-crystal graphene by suppressing evaporative loss of Cu during low pressure chemical vapor deposition. Adv. Mater. \textbf{25}, 2062-2065, (2013).


\bibitem{Li09}
X. Li, W. Cai, J. An, S. Kim, J. Nah, D. Yang, R. Piner, A. Velamakanni, I. Jung, E. Tutuc, S. K. Banerjee, L. Colombo, R. S. Ruoff, Large-area synthesis of high-quality and uniform graphene films on copper foils. Science \textbf{324}, 1312, (2009).


\bibitem{Bae10}
S. Bae, H. Kim, Y. Lee, X. Xu, J. S. Park, Y. Zheng, J. Balakrishnan, T. Lei, H. R. Kim, Y. I. Song, Y. J. Kim, K. S. Kim, B. Ozyilmaz, J. H. Ahn, B. H. Hong, S. Iijima,
Roll-to-roll production of 30-inch graphene films for transparent electrodes, Nat. Nanotechnol. \textbf{5}, 574, (2010).


\bibitem{Li11}
Xuesong Li, Carl W. Magnuson, Archana Venugopal, Rudolf M. Tromp, James B. Hannon, Eric M. Vogel, Luigi Colombo, and Rodney S. Ruoff, Large-Area Graphene Single Crystals Grown by Low-Pressure Chemical Vapor Deposition of Methane on Copper. J. Am. Chem. Soc. \textbf{133}, 2816, (2011).



\bibitem{Ban15}
L. Banszerus, M. Schmitz, S. Engels, J. Dauber, M. Oellers, F. Haupt, K. Watanabe, T. Taniguchi, B. Beschoten, C. Stampfer, Ultrahigh-mobility graphene devices from chemical vapor deposition on reusable copper. Sci. Adv. \textbf{1}, e1500222, (2015).


 \bibitem{Suk11}
 J. W. Suk, A. Kitt, C. W. Magnuson, Y. Hao, S. Ahmed, J. An, A. K. Swan, B. B. Goldberg, R. S. Ruoff,
 Transfer of CVD-grown monolayer graphene onto arbitrary substrates. ACS Nano \textbf{5}, 6916, (2011).

  \bibitem{Piz15}
F. Pizzocchero, B. Jessen, P. Whelan, N. Kostesha, S. Lee, J. Buron, I. Petrushina, M. Larsen, P. Greenwood, W. Cha, K. Teo, P. Jepsen, J. Hone, P. B\o
ggild, T. Booth, Non-destructive electrochemical graphene transfer from reusable thin-film catalysts.  Carbon \textbf{85}, 397-405, (2015).



\bibitem{Pet12}
N. Petrone, C. R. Dean, I. Meric, A. M. van der Zande, P. Y. Huang, L.Wang, D. Muller, K. L. Shepard, and J. Hone, Chemical vapor deposition-derived graphene with electrical performance of exfoliated graphene. Nano Lett. \textbf{12}, 2751, (2012).


 \bibitem{Wan15}
Z. Wang, D.-K. Ki, H. Chen, H. Berger, A. H. MacDonald and A. F. Morpurgo, Strong interface-induced spin-orbit interaction in graphene on WS2. Nature Communications \textbf{6}, 8339, (2015).

 \bibitem{Avs14}
 A. Avsar, J. Y. Tan, T. Taychatanapat, J. Balakrishnan, G. K. W. Koon, Y. Yeo, J. Lahiri, A. Carvalho, A. S. Rodin, E.C.T. O'Farrell, G. Eda, A. H. Castro Neto and B. \"Ozyilmaz, Spin-orbit proximity effect in graphene. Nature Commun. \textbf{5}, 4875, (2014).

\bibitem{Brit13}
L. Britnell, R. M. Ribeiro, A. Eckmann, R. Jalil, B.D. Belle, A. Mishchenko, Y.-J. Kim,  R.V. Gorbachev, T. Georgiou, S.V. Morozov, A.N. Grigorenko, A. K. Geim, C. Casiraghi, A.H. Castro Neto, K.S. Novoselov, Strong Light-Matter Interactions in Heterostructures of Atomically Thin Films. Science \textbf{340}, 6138, {2013}.

 \bibitem{Las13}
M. R. Laskar, L. Ma, S. Kannappan, P. S. Park, S. Krishnamoorthy, D. N. Nath, W. Lu, Y. Wu and S. Rajan, Large area single crystal (0001) oriented MoS$_2$. Appl. Phys. Lett. \textbf{102}, 252108, (2013).

\bibitem{Eic15}
S. M. Eichfeld, L. Hossain, Y.-C. Lin, A. F. Piasecki, B. Kupp, A. G. Birdwell, R. A. Burke, N. Lu, X. Peng, J. Li, A. Azcatl, S. McDonnell, R. M. Wallace, M. J. Kim, T. S. Mayer, J. M. Redwing, and J. A. Robinson, Highly Scalable, Atomically Thin WSe$_2$ Grown via Metal-Organic Chemical Vapor Deposition. ACS Nano \textbf{9}, 2080-2087, (2015).

\bibitem{Che16}
J. Chen, W. Tang, B. Tian, B. Liu, X. Zhao, Y. Liu, T. Ren, W. Liu, D. Geng, H. Y. Jeong, H. S. Shin, W. Zhou, K. P. Loh, Chemical Vapor Deposition of High-Quality Large-Sized MoS$_2$ Crystals on Silicon Dioxide Substrates. Adv. Sci., \textbf{3} 1600033, (2016).



\bibitem{Lu12}
A. Y. Lu, S.-Y. Wei, C.-Y. Wu, Y. Hernandez, T.-Y. Chen, T.-H. Liu, C.-W. Pao, F.-R. Chen, L.-J. Li and Z.-Y. Juang, Decoupling of CVD graphene by controlled oxidation of recrystallized Cu. RSC Adv. \textbf{2}, 3008-3013, (2012).


 \bibitem{Fer06}
  A. C. Ferrari, J. C. Meyer, V. Scardaci, C. Casiraghi, M. Lazzeri, F. Mauri, S. Piscanec, D. Jiang, K. S. Novoselov, S. Roth, A. K. Geim,
  Raman spectrum of graphene and graphene layers. Phys. Rev. Lett. \textbf{97}, 187401, (2006).


 \bibitem{Gra07}
 D. Graf, F. Molitor, K. Ensslin, C. Stampfer, A. Jungen, C. Hierold, L. Wirtz,
 Spatially resolved Raman spectroscopy of single- and few-layer graphene. Nano Lett. \textbf{7}, 238-242, (2007).


 \bibitem{For13}
 F. Forster, A. Molina-Sanchez, S. Engels, A. Epping, K. Watanabe, T. Taniguchi, L. Wirtz, C. Stampfer,
  Dielectric screening of the Kohn anomaly of graphene on hexagonal boron nitride. Phys. Rev. B \textbf{88}, 085419, (2013).


 \bibitem{Neu14}
 C. Neumann, S. Reichardt, P. Venezuela, M. Dr\"ogeler, L. Banszerus, M. Schmitz, K. Watanabe, T. Taniguchi, F. Mauri, B. Beschoten, S. V. Rotkin, C. Stampfer,
 Raman spectroscopy as probe of nanometer-scale strain variations in graphene. Nature Commun. \textbf{6}, 8429, (2015).

 \bibitem{Lee12}
 J. E. Lee, G. Ahn, J. Shim, Y. S. Lee, S. Ryu, Optical separation of mechanical strain from charge doping in graphene. Nat. Commun. \textbf{3}, 1024, (2012).



\bibitem{Ton13}
P. Tonndorf, R. Schmidt, P. B\"ottger, X. Zhang, J. B\"orner, A. Liebig, M. Albrecht, C. Kloc, O. Gordan, D. R. T. Zahn, S. Michaelis de Vasconcellos, and R. Bratschitsch, Photoluminescence emission and Raman response of monolayer MoS$_2$, MoSe$_2$, and WSe$_2$. Opt. Express \textbf{21}, 4908-4916, (2013).


\bibitem{Sah13}
H. Sahin, S. Tongay, S. Horzum, W. Fan, J. Zhou, J. Li, J. Wu, and F. M. Peeters, Anomalous Raman spectra and thickness-dependent electronic properties of WSe$_2$. Phys. Rev. B \textbf{87}, 165409, (2013).

   \bibitem{Cuo14}
   N. J. G. Couto, D. Costanzo, S. Engels, D.-K. Ki, K. Watanabe, T. Taniguchi, C. Stampfer, F. Guinea, and A. F. Morpurgo, Random Strain Fluctuations as Dominant Disorder Source for High-Quality On-Substrate Graphene Devices. Phys. Rev. X \textbf{4}, 041019, (2014).

\bibitem{Ban17}
L. Banszerus, H. Janssen, M. Otto, A. Epping, T. Taniguchi, K. Watanabe, B. Beschoten, D. Neumaier and C. Stampfer, Identifying suitable substrates for high-quality graphene-based heterostructures. 2D Mater. \textbf{4}, 025030, (2017).


\bibitem{Tao12}
L. Tao, J. Lee, H. Chou, M. Holt, R. S. Ruoff, and D. Akinwande, Synthesis of High Quality Monolayer Graphene at Reduced Temperature on Hydrogen-Enriched Evaporated Copper (111) Films. ACS Nano \textbf{6}, 2319-2325, (2012).


\bibitem{Gao10}
L. Gao, J. R. Guest and N. P. Guisinger, Epitaxial Graphene on Cu(111). Nano Lett. \textbf{10}, 3512-3516, (2010).




\bibitem{Hua12}
 B. Hua, H. Agoa, Y. Itob, K. Kawaharaa, M. Tsujia, E. Magomec, K. Sumitanic, N. Mizutab, K.-i. Ikedab, S. Mizunob, Epitaxial growth of large-area single-layer graphene over Cu(111)/sapphire by atmospheric pressure CVD. Carbon \textbf{50}, 57-65, (2012).


\bibitem{Zha05}
Y. Zhang, Y.W. Tan, H. L. Stormer, and P. Kim, Experimental observation of the quantum Hall effect and Berry's phase in graphene. Nature \textbf{438}, 201-204, (2005).

\bibitem{ref}
X. Cui, G.-H. Lee, Y. D. Kim, G. Arefe, P. Y. Huang, C.-H. Lee, D. A. Chenet, X. Zhang, L. Wang, F. Ye, F. Pizzocchero, B. S. Jessen, K. Watanabe, T. Taniguchi, D. A. Muller, T. Low, P. Kim and J. Hone, Multi-terminal transport measurements of MoS$_2$ using a van der Waals heterostructure device platform. Nat. Nanotechnol. \textbf{10}, 534-540, (2015).


\end{thebibliography}
%

\end{document}